\documentstyle[twocolumn,floats,aps,prb,epsf]{revtex}
\draft
\title{Muon Spin Relaxation and Nonmagnetic Kondo State in PrInAg$\bf{_2}$}
\author{D. E. MacLaughlin}
\address{MS K764, Los Alamos National Laboratory, Los Alamos, New Mexico 87545 
\\ Department of Physics, University of California, Riverside, California 
92521-0413}
\author{R. H. Heffner}
\address{MS K764, Los Alamos National Laboratory, Los Alamos, New Mexico 87545}
\author{G. J. Nieuwenhuys}
\address{Kamerlingh Onnes Laboratorium, Leiden University, \\ 
2500 RA Leiden, The Netherlands}
\author{P. C. Canfield}
\address{Ames Laboratory and Department of Physics and Astronomy \\ 
Iowa State University, Ames, Iowa 50011}
\author{A. Amato, C. Baines}
\address{Paul Scherrer Institute, CH-5232 Villigen, Switzerland}
\author{A. Schenck}
\address{Institut f\"ur Teilchenphysik, Eidgen\"ossiche Technische 
Hochschule-Z\"urich, \\ CH-5232 Villigen, Switzerland}
\author{G. M. Luke, Y. Fudamoto, Y. J. Uemura}
\address{Department of Physics, Columbia University, New York, New York 10027}

\author{\small(September 16, 1998)}

\address{\parbox{14cm}{\bigskip\rm\small			
Muon spin relaxation experiments have been carried out in the Kondo 
compound~PrInAg$_2$. The zero-field muon relaxation rate is found to be 
independent of temperature between 0.1 and 10 K, which rules out a magnetic 
origin (spin freezing or a conventional Kondo effect) for the 
previously-observed specific heat anomaly at $\sim$0.5 K\@. The low-temperature 
muon relaxation is quantitatively consistent with nuclear magnetism including 
hyperfine enhancement of the $^{141}$Pr nuclear moment. This is strong evidence 
against a Pr$^{3+}$ electronic magnetic moment, and for the $\Gamma_3$ 
crystalline-electric-field-split ground state required for a nonmagnetic route 
to heavy-electron behavior. The data imply the existence of an exchange 
interaction between neighboring Pr$^{3+}$ ions of the order of 0.2 K in 
temperature units, which should be taken into account in a complete theory of a 
nonmagnetic Kondo effect in PrInAg$_2$.
\\[6pt] PACS numbers : 71.27.+a, 72.15.Qm, 76.75.+i, 75.30.Mb}}
									
\begin{document} \maketitle					

\thispagestyle{myheadings} \markright{{\small Submitted to {\em Physical Review B}}\hfill {\small LA-UR-98-3265\quad p.}\hspace{1mm}} 

\section{Introduction}
In a seminal paper, Yatskar {\em et al.\/}\cite{YBMC96} reported evidence for 
unconventional heavy-fermion behavior in the praseodymium-based 
intermetallic~PrInAg$_2$. This compound is one of only a handful of Pr-based 
materials which exhibit heavy-fermion or Kondo-like properties. Specific heat, 
magnetic susceptibility, and neutron scattering experiments\cite{GPSM84} 
indicate a non-Kramers doublet ($\Gamma_3$) ground state due to 
crystalline-electric-field (CEF) splitting of the Pr$^{3+}$ $^1H_4$ term. The 
$\Gamma_3$ state is {\em nonmagnetic\/}, i.e., there are no matrix elements of 
the magnetic moment operator within its doubly degenerate manifold. A 
nonmagnetic ground state would make the heavy-fermion-like specific heat anomaly 
found below 1 K and the enormous low-temperature Sommerfeld specific heat 
coefficient~$\gamma(T) \approx 6.5$ J\,mole$^{-1}$\,K$^{-2}$ quite unexpected, 
and suggests that PrInAg$_2$ may be a system in which an unusual nonmagnetic 
path to heavy-fermion behavior is realized.\cite{YBMC96}\, But such a scenario 
depends crucially on the nonmagnetic nature of the ground state.

This paper reports two results of muon-spin-relaxation ($\mu$SR) experiments in 
PrInAg$_2$ which support the conclusion of Yatskar {\em et al.\/}\cite{YBMC96} 
that the Kondo effect in PrInAg$_2$ is nonmagnetic in origin. First, we observe 
no temperature dependence of the muon relaxation rate at low temperatures, 
contrary to what would be expected if the specific heat anomaly involved 
magnetic degrees of freedom. Second, the muon relaxation behavior indicates that 
the CEF ground state in PrInAg$_2$ is in fact nonmagnetic, since the 
low-temperature muon relaxation can only be understood in terms of strong {\em 
hyperfine enhancement\/} of the $^{141}$Pr nuclear magnetism. Hyperfine 
enhancement is an effect of the hyperfine coupling between the nucleus and the 
Van Vleck susceptibility of $f$ electrons of a non-Kramers $f$ ion in a 
nonmagnetic ground state,\cite{Alts67} and only occurs when the Pr$^{3+}$ CEF 
ground state is nonmagnetic.\cite{Tepl77}\, It has been used to attain very low 
temperatures by nuclear demagnetization of singlet-ground-state Pr-based 
intermetallics.\cite{Andr78} 

Both of our results confirm the unusual nonmagnetic correlated-electron behavior 
of PrInAg$_2$, using for the first time a microscopic probe of the electronic 
structure. We argue below that $\mu^+$ relaxation is dominated by dipolar 
coupling to nearby $^{115}$In and $^{141}$Pr nuclear magnetic moments (Ag 
nuclear moments are negligible in comparison); the Pr$^{3+}$ wave function 
enters only in the hyperfine enhancement of the $^{141}$Pr nuclear magnetism. 
The fact that no direct Pr$^{3+}$ $f$-ion magnetism is observed in PrInAg$_2$ is 
strong evidence for a nonmagnetic Pr$^{3+}$ CEF ground state, and the fact that 
the specific heat anomaly of Yatskar {\em et al.\/}\ corresponds to a molar 
entropy of $R \ln 2$ indicates that this ground state is a $\Gamma_3$ doublet.

The remainder of this introduction contains three brief pedagogical sections: a 
description of the theoretical basis for a nonmagnetic Kondo effect 
(Sect.~\ref{sect:QKE}), an introduction to the elements of the $\mu$SR technique 
used in this study (Section~\ref{sect:muSR}), and a review of the important 
aspects of hyperfine enhancement (Sect.~\ref{sect:hfenh})\@. In 
Sect.~\ref{sect:results} we describe our experimental results in PrInAg$_2$, 
which include the temperature dependence of the zero-field $\mu$SR relaxation 
and the longitudinal field dependence of the relaxation at low temperatures. The 
implications of these results for the nature of the low-temperature state of 
PrInAg$_2$ are discussed in Sect.~\ref{sect:disc}, where it is concluded that 
(a)~the observed $\mu$SR behavior is due to nuclear magnetism rather than a 
magnetic Pr$^{3+}$ ground state, i.e., the ground state is nonmagnetic; and 
(b)~the effect of the muon electric charge on its environment does not 
invalidate the analysis which leads to this conclusion. We summarize our results 
in Sect.~\ref{sect:concl}.

\subsection{Nonmagnetic Kondo effect} \label{sect:QKE}
The only nonmagnetic mechanism for Kondo behavior proposed to date is the 
two-channel quadrupolar Kondo effect (QKE) of Cox.\cite{Cox87}\, In this 
picture, which was developed to explain the unexpected lack of field dependence 
of heavy-fermion properties in uranium-based compounds, correlated-electron 
behavior occurs when a non-Kramers $f$ ion such as Pr$^{3+}$ possesses a 
nonmagnetic multiplet ground state. The fluctuating electric quadrupole moment 
of the ground state scatters conduction electrons, analogous to spin-fluctuation 
scattering in the usual Kondo effect. An important difference between the two 
effects is that in the QKE there are two conduction-electron channels (spin-up 
and spin-down); since spin plays no role in the nonmagnetic scattering, the spin 
directions serve only as labels. The QKE is therefore one of a class of {\em 
multi-channel\/} Kondo effects\cite{SaSc89,CoJa96} for which the low-temperature 
behavior is that of a ``non-Fermi liquid'' with unusual properties, e.g., 
logarithmic divergence of $\gamma(T)$ and nonzero residual entropy~$S(T{=}0) = \mbox{$\frac{1}{2}$}R\ln 2$. 

In its original form the theory of the QKE considers isolated impurities only, 
and to our knowledge no treatment of a lattice of nonmagnetic QKE $f$ ions has 
appeared.  In particular, it is apparently not known whether the 
non-Fermi-liquid behavior of the impurity problem survives in the lattice. 
Although Yatskar {\em et al.\/}\cite{YBMC96} observed an uncharacteristic 
temperature dependence of the low-temperature electrical resistivity in 
PrInAg$_2$, they found a substantially temperature-independent $\gamma(T)$ below 
$\sim$0.2 K and no evidence for residual entropy. Thus it is unclear whether or 
not PrInAg$_2$ is a Fermi liquid. 

\subsection{Zero- and low-field muon spin relaxation} \label{sect:muSR}
$\mu$SR is a sensitive local probe of static and dynamic magnetism in 
solids.\cite{Sche85}\, Spin-polarized positive muons ($\mu^+$) are implanted 
into the sample, and the subsequent decay of the $\mu^+$ spin polarization is 
monitored in time by measuring the asymmetry in the numbers of muon decay 
positrons emitted parallel and antiparallel to the $\mu^+$ spin direction. The 
resulting relaxation function~$G(t)$ is analogous to the free induction signal 
in a pulsed nuclear magnetic resonance (NMR) experiment. It is straightforward 
to carry out $\mu$SR experiments in zero or weak applied magnetic fields, which 
is not the case for NMR. 

The shape and duration of $G(t)$ is controlled by the local magnetic fields at 
the $\mu^+$ sites due to their magnetic environments. There are two kinds of 
effects. Relaxation by {\em static\/} local fields reflects a spatial 
distribution of $\mu^+$ Larmor precession frequencies and hence of the local 
fields. The decay of $G(t)$ is then due to loss of phase coherence between 
precessing $\mu^+$ spins, and the relaxation time is of the order of the inverse 
of the spread in Larmor frequencies. If the $\mu^+$ local field distribution is 
due to randomly-oriented neighboring magnetic dipole moments (nuclear or 
electronic), the Central Limit Theorem suggests a Gaussian field distribution if 
more than a few moments contribute, in which case $G(t)$ is also Gaussian. 
Fields due to randomly-oriented nuclear dipolar moments, which usually do not 
reorient on the time scale of $\mu$SR experiments,\cite{Prrelax} often give rise 
to static relaxation. $\mu$SR is also a very good test for static magnetism, 
with or without long-range order, with a sensitivity ${\sim}10^{-3} \mu_B$, 
since dipolar fields from such small moments produce observable static 
relaxation. {\em Dynamic\/} (fluctuating) $\mu^+$ local fields lead to 
spin-lattice relaxation, as in NMR, which is a measure of the spectral density 
of the fluctuations at low frequencies. For dynamic relaxation $G(t)$ is often 
but not always exponential.
 
Static and dynamic relaxation mechanisms can be distinguished by $\mu$SR 
experiments in a longitudinal field~${\bf H}_L$ (i.e., a field parallel to the 
$\mu^+$ spin direction) much larger in magnitude than a typical local 
field~${\bf H}_{\rm loc}$. This produces a resultant field~${\bf H}_L + {\bf 
H}_{\rm loc}$ essentially in the direction of the applied field and hence of the 
$\mu^+$ spin. Then the muons do not precess substantially, and if ${\bf H}_{\rm 
loc}$ is static their spin polarization is maintained indefinitely. This 
procedure is known as ``decoupling'' of the $\mu^+$ spin from the static local 
fields. If on the other hand the relaxation is dynamic, then it is usually much 
less affected by the relatively weak applied field (typically $H_L \lesssim 100$ 
Oe). The expected field for decoupling is a few times the spread~$\Delta H_{\rm 
loc}$ in local fields, which can be estimated self-consistently by assuming that 
the relaxation is static. In this case the observed relaxation rate gives the 
spread~$\sigma$ of $\mu^+$ precession frequencies, so that
\begin{equation} 
\Delta H_{\rm loc} = \sigma /\gamma_\mu \,, 
\label{eq:DeltaH}
\end{equation}
where $\gamma_\mu$ is the $\mu^+$ gyromagnetic ratio.

Zero-field and low-field $\mu$SR (ZF- and LF-$\mu$SR) relaxation data are often 
analyzed using the Kubo-Toyabe (K-T) model,\cite{KuTo67} in which the shape of 
the relaxation function and its rate of decay are functions of the parameters 
which characterize the local field distribution and dynamics (i.e., $\sigma$ and 
the fluctuation rate~$\nu$ of the local field), and also the conditions of 
measurement (i.e., the val\-ue of the applied field and its orientation relative 
to the $\mu^+$ spin direction). 

\subsection{Hyperfine-enhanced nuclear magnetism} \label{sect:hfenh}
The best-known hyperfine-enhancement effect is the enhancement of the applied 
field at the nuclear site by a factor~$1 + K$ (Ref.~\onlinecite{Alts67}), with
\begin{equation}
K = a_{\rm hf}\,\chi_{\rm VV} \,.
\label{eq:K}
\end{equation}
Here $a_{\rm hf}$ is the $f$ atomic hyperfine coupling constant, expressed in 
units of mole\,emu$^{-1}$, and $\chi_{\rm VV}$ is the (molar) Van Vleck 
susceptibility of the $f$ ions. The factor~$K$ is the usual paramagnetic NMR 
frequency shift (Knight shift in metals), due in this case to $\chi_{\rm VV}$. 
The latter is given approximately by
\begin{equation}
\chi_{\rm VV} \approx \frac{\cal C}{(\Delta_{\rm CEF}/k_B)} \,,
\label{eq:chiVV}
\end{equation}
where $\cal C$ is the $f$-ion Curie constant and $\Delta_{\rm CEF}$ is the 
excitation energy of the lowest CEF magnetic excited state. For typical 
Pr$^{3+}$ splittings $\Delta_{\rm CEF} = 10$--100 K, so that $\chi_{\rm VV} = 
0.01$--0.1 emu\,mole$^{-1}$. With $a_{\rm hf}({\rm Pr}^{3+}) = 187.7$ 
mole\,emu$^{-1}$ (Ref.~\onlinecite{AnDa77}), one finds $K = 2$--20. These 
considerable field increases are exploited in hyperfine-enhanced nuclear 
cooling.\cite{AnBu68}

Other effects of hyperfine enhancement include the following:\cite{Blea73}
\begin{itemize}

\item The $f$ electrons are polarized by the nuclear magnetic dipole moment via 
a Van Vleck-like response, leading to an enhanced effective nuclear 
moment~$\mu_{\rm nuc}^{\rm{(enh)}} = (1 + K)\mu_{\rm nuc}^{\rm (bare)}$. The 
nuclear moment itself is of course unchanged; the term~``hyperfine 
enhancement'', used here in much the same sense as ``many-body enhancement of 
the electron mass'' in heavy-fermion systems, refers to the interaction of the 
effective (nuclear + $f$-electron) moment with its magnetic environment.

\item The electronic exchange coupling between neighboring $f$ ions mediates an 
indirect exchange interaction between nuclear spins, with exchange 
constant~${\cal J}_{\rm nuc}$ given by
\begin{equation}
{\cal J}_{\rm nuc} = \left( \frac{\gamma_{\rm nuc}\hbar}{g_J\mu_B} 
\right)^{\!\!2} K^2 {\cal J}_{\rm el} \,;
\label{eq:Jnuc} 
\end{equation}
here $\gamma_{\rm nuc}$ is the nuclear gyromagnetic moment and ${\cal J}_{\rm 
el}$ is the electronic exchange constant.

\end{itemize}
As Bleaney\cite{Blea73} puts it, ``\dots we are dealing with nuclear lambs in 
electronic wolves' clothing.''

Hyperfine enhancement phenomena were first observed in singlet ground-state 
$f$-ion compounds, but are also expected for $f$ ions with nonmagnetic multiplet 
ground states. Such effects do not occur at temperatures $\gtrsim \Delta_{\rm 
CEF}/k_B$ (Ref.~\onlinecite{Tepl77}) or if the ground state of the $f$ ion is 
magnetic. In both of these circumstances the $f$-electron polarization induced 
by the nucleus is lost in the much larger electronic magnetic moment of the $f$ 
ion, and hyperfine-enhancement phenomena are obliterated.

\section{Experimental results} \label{sect:results}
The sample of PrInAg$_2$ was prepared as described previously.\cite{YBMC96}\, 
$\mu$SR experiments were carried out at the $\pi$M3 beam line of the Paul 
Scherrer Institute (PSI), Villigen, Switzerland, using the General Purpose 
Spectrometer (GPS) and Low Temperature Facility (LTF), and at the M20 beam line 
at TRIUMF, Vancouver, Canada. ZF-$\mu$SR data were obtained over the temperature 
range 0.1--100 K, and the dependence of the relaxation function on longitudinal 
field was studied at a temperature of 0.7 K\@. We describe these results below, 
and discuss their implications in Sect.~\ref{sect:disc}.

\subsection{ZF muon relaxation between 0.1 K and 100 K} \label{sect:ZFrelax}
As a preliminary characterization of the ZF relaxation function we fit the 
relaxation data to a ``power exponential'' function 
\[ G(t) = \exp\left[-(\Lambda t)^\beta\right] \,,  \]
where $\Lambda$ is a generalized relaxation rate and the exponent~$\beta$ 
interpolates between exponential ($\beta = 1$) and Gaussian ($\beta = 2$) limits. 
This fit function and parameters give a rough indication of the behavior of the 
relaxation, i.e., whether as discussed in Sect.~\ref{sect:muSR} it is dynamic 
(exponential), static (Gaussian), or an intermediate case.

\begin{figure}%
[ht] \epsfxsize 3.4in	\epsfbox{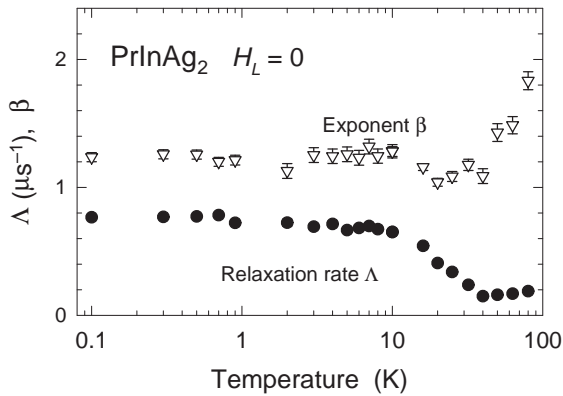}			
\caption{Temperature dependence of zero-field $\mu^+$ relaxation rate~$\Lambda$ 
(circles) and exponent~$\beta$ (triangles) in PrInAg$_2$ from fits of a ``power 
exponential'' function~$G(t) = \exp[-(\Lambda t)^\beta]$ to ZF-$\mu$SR 
relaxation data.}
\label{fig:ZFtempdep}
\end{figure}

The temperature dependence of $\Lambda$ and $\beta$ is given in 
Fig.~\ref{fig:ZFtempdep}. 
Both parameters are essentially independent of temperature from 0.1 K to 
$\sim$10\ K\@. At low temperatures $\Lambda \approx 0.8\ \mu{\rm s}^{-1}$; we 
shall see that this is much larger than expected from $^{115}$In and ``bare'' 
$^{141}$Pr nuclear dipole fields. The low-temperature val\-ue $\beta \approx 
1.2$ indicates that the relaxation is nearly but not quite exponential.

Above 10 K both $\Lambda$ and $\beta$ vary with temperature. A decrease of 
$\Lambda$ to $0.18 \pm 0.02\ \mu{\rm s}^{-1}$ occurs between $\sim$10 K and 
$\sim$50 K\@, and above 50 K $\beta$ increases to ${\lesssim}2$ at 80 K\@. 
This suggests a crossover to static relaxation at high temperatures. Although 
LF decoupling experiments have not been carried out above 0.7 K, the Gaussian 
shape and slow rate of the high-temperature muon relaxation function are 
consistent with nuclear dipolar relaxation in the static limit.

\subsection{ZF and LF muon relaxation at 0.7 K}
\begin{figure}%
[ht] \epsfxsize 3.4in	\epsfbox{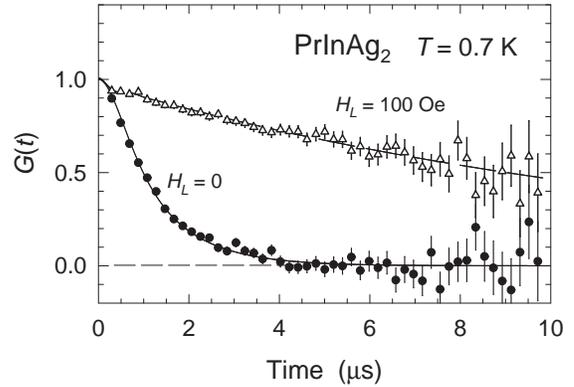}			
\caption{Dependence of $\mu^+$ relaxation function~$G(t)$ on longitudinal 
applied field~$H_{L}$ in PrInAg$_2$, $T = 0.7$ K\@. The relaxation in 100 Oe 
(triangles) is much faster than expected if the zero-field relaxation (circles) 
were due to a static distribution of local fields. Curves: fits to dynamic K-T 
model (Refs.~\protect\onlinecite{KuTo67} and \protect\onlinecite{HUIN79}) for 
$H_L = 0$ (solid curve) and $H_L = 100$ Oe (dashed curve).}
\label{fig:LFdep}
\end{figure}

Figure~\ref{fig:LFdep} shows the experimental $\mu^+$ relaxation 
functions~$G(t)$ for $T = 0.7$ K in zero field and in a longitudinal applied 
field~$H_L = 100$ Oe. 
If the relaxation were due to a distribution of static $\mu^+$ local fields, the 
observed ZF relaxation rate leads to an estimate of $\sim$10 Oe for the spread 
of these local fields. Then a longitudinal field of 100 Oe should completely 
decouple the local fields and there should be no relaxation. But it can be seen 
from Fig.~\ref{fig:LFdep} that although the relaxation rate is reduced for $H_L 
= 100$ Oe it remains appreciable (${\sim}0.07\ \mu{\rm s}^{-1}$). This strongly 
suggests that dynamic relaxation is involved. The curves in Fig.~\ref{fig:LFdep} 
are fits to the dynamic K-T model as described below in 
Sect.~\ref{sect:ltrelax}.

\section{Discussion} \label{sect:disc}
In this section we first consider the implications of the 
temperature-independent ZF relaxation between 0.1 K and 10 K, using the rough 
``power exponential'' analysis of Sect.~\ref{sect:ZFrelax} and the data of 
Fig.~\ref{fig:ZFtempdep}. We then carry out a K-T analysis of the data of 
Fig.~\ref{fig:LFdep} and obtain the rms width and fluctuation rate of the local 
field at 0.7 K\@. Finally, we show that (a)~the rms widths at 0.7 K and 100 K 
are quantitatively explained by nuclear magnetism (including $^{141}$Pr 
hyperfine enhancement at low temperatures) alone, so that no Pr$^{3+}$ 
electronic magnetic moment need be invoked, and (b)~the $^{141}$Pr fluctuation 
rate at 0.7 K is consistent with the indirect nuclear exchange mechanism 
described above in Sect.~\ref{sect:hfenh}.

\subsection{Zero-field muon relaxation rate at low temperatures} 
\label{sect:ZFmuSR}
The data of Fig.~\ref{fig:ZFtempdep} put an upper bound of ${\sim}0.05\ \mu{\rm 
s}^{-1}$ on any change of $\Lambda$ between 0.1 K and 10 K\@. This result has 
important implications for the two possible explanations of the low-temperature 
anomalies of Yatskar {\em et al.\/}\cite{YBMC96} which involve hypothetical 
Pr$^{3+}$ magnetic moments: (a)~weak-moment static magnetism, and 
(b)~``conventional'' Kondo spin fluctuations.

\paragraph{Weak-moment static magnetism due to spin freezing.} We show below in 
Sect.~\ref{sect:Prnuc} that in the absence of hyperfine-enhancement effects the 
(quasistatic) nuclear contribution to the ZF muon relaxation rate is 
considerably less than $0.8\ \mu{\rm s}^{-1}$. Thus the persistence of such 
rapid relaxation to 10 K and above is already an indication that its cause is 
not related to the onset of static magnetism below 0.5--1 K, which would be 
required to explain the specific heat anomaly by such spin freezing. We 
nevertheless take the observed relaxation rate~$\Lambda(T)$ to be due to a 
temperature-independent component~$\Lambda_0$ and static relaxation and a 
temperature-dependent component~$\lambda(T)$ due to the spin freezing; the 
latter presumably exhibits an anomaly in the neighborhood of the freezing 
temperature. Assuming conservatively that these components add in quadrature we 
have
\[ \Lambda(T) = \left[ \Lambda_0^2 + \lambda(T)^2 \right]^{1/2} \,. \]
With $\Lambda_0  \approx 0.8\ \mu{\rm s}^{-1}$ and $\Lambda(T) - \Lambda_0 
\lesssim 0.05\ \mu{\rm s}^{-1}$ we obtain
\[ \lambda(T) \lesssim 0.29\ \mu{\rm s}^{-1} \,, \]
which yields an upper bound on the frozen Pr$^{3+}$ magnetic moment of order
\[ \mu({\rm Pr}^{3+}) \approx 3 \times 10^{-3}\mu_B \,. \]
Static magnetism is therefore ruled out at this level. If $\Lambda_0$ and 
$\lambda(T)$ were assumed to add linearly the upper limit on $\lambda(T)$ would 
be much smaller: $\lambda(T) \lesssim 0.05\ \mu{\rm s}^{-1}$. Thus Pr$^{3+}$ 
spin freezing fails to account for our results, and in addition leaves 
unexplained the fast and temperature-independent $\mu^+$ relaxation above 
$\sim$1 K.

\paragraph{Conventional (spin) Kondo physics.} For temperatures $\lesssim T_K$ a 
Kondo spin fluctuates at a rate~$\nu \approx k_BT_K/\hbar$, which is 
${\sim}10^{11}\ {\rm s}^{-1}$ for $T_K \approx 1$ K\@. In the neighborhood of 
$T_K$ the $\mu^+$ relaxation rate~$T_1^{-1}$ is a maximum given by
\[ (T_1^{-1})_{\rm max} \approx 2\sigma^2/\nu\,, \]
where $\sigma$ is the spread of $\mu^+$ frequencies due to Pr$^{3+}$ dipole 
moments. We estimate $\sigma$ using Eq.~(\ref{eq:DeltaH}), with the 
val\-ue~$\Delta H_{\rm loc} = 1.10\ {\rm kOe}/\mu_B$ calculated via a lattice 
sum over Pr sites assuming uncorrelated Pr-moment fluctuations. This gives
\[  (T_1^{-1})_{\rm max} \approx 0.2\ \mu{\rm s}^{-1} \, \]
for a Pr$^{3+}$ moment of the order of 1 $\mu_B$.\cite{Prmoment}\, We expect 
this rate for $T \gtrsim T_K$, with a crossover to a Korringa law
\begin{equation}
T_1^{-1} \approx (T_1^{-1})_{\rm max} \left(\frac{T}{T_K}\right) 
\label{eq:T1K}
\end{equation}
(i.e., a considerable decrease of $T_1^{-1}$) for $T < T_K$. 

As discussed above in connection with the possibility of static magnetism, a 
change of this order of magnitude is in fact not seen: the observed 
rate~$\Lambda \approx 0.8\ \mu{\rm s}^{-1}$ is constant down to $\sim$0.1 K 
(Fig.~\ref{fig:ZFtempdep}), whereas from Eq.~(\ref{eq:T1K}) and $T_K \approx 1$ 
K (Ref.~\onlinecite{YBMC96}) one expects a decrease of $T_1^{-1}$ at 0.1 K to 
$\sim$10\% of its val\-ue at 1 K\@. It might be argued that the Korringa rate is 
masked by a temperature-independent rate~$\Lambda_0$, as in the above discussion 
of spin freezing, in which case the minimum observable change of $0.29\ \mu{\rm 
s}^{-1}$ derived above would apply and a Korringa-like change of ${\sim}0.2\ 
\mu{\rm s}^{-1}$ could not be ruled out by the data. But in the ``conventional 
Kondo'' scenario, as in the spin-freezing picture, there is no mechanism for a 
temperature-independent rate as large as $\Lambda_0$. We therefore conclude 
that the experimental results are not consistent with conventional Kondo 
behavior. 

\subsection{ZF and LF muon relaxation at 0.7 K} \label{sect:ltrelax}
We compare the relaxation functions at $T = 0.7$ K (Fig.~\ref{fig:LFdep}) with 
the K-T model in the ``strong-collision'' approximation of Hayano {\em et 
al.\/}\cite{HUIN79}\, This approximation takes the fluctuation to be of the form 
of uncorrelated sudden jumps of the local field. We make it as a matter of 
convenience, since it is easier to treat numerically than the Gaussian-Markovian 
process originally described by Kubo and Toyabe\cite{KuTo67} and the 
quantitative differences are small.

The model assumes that the $\mu^+$ local field~${\bf H}_{\rm loc}$ is 
distributed in magnitude and direction, with each Cartesian component 
distributed around zero with rms val\-ue $\sigma/\gamma_\mu$, and that ${\bf 
H}_{\rm loc}$ fluctuates randomly in time with fluctuation rate $\nu$. In the 
following we refer to $\sigma$ as the static relaxation rate. In zero field a 
crossover occurs with increasing fluctuation rate from the ``quasistatic'' 
regime ($\nu \ll \sigma$) to the ``motionally-narrowed'' regime ($\nu \gg 
\sigma$). For $\nu \gtrsim \sigma$ the main dependence of the ZF relaxation 
function is on the combination~$\sigma^2/\nu$; the dependence on $\sigma$ and 
$\nu$ separately is weaker (and vanishes in the extreme narrowing limit~$\nu \gg 
\sigma$). Thus the ZF data alone do not suffice to determine both parameters, 
and the added constraint provided by the applied field is essential. This 
circumstance also prohibits a meaningful comparison of the ZF muon relaxation at 
other temperatures with the K-T model.

The best agreement with both ZF and LF relaxation data, given by the curves in 
Fig.~\ref{fig:LFdep}, is obtained for $\sigma = 1.15 \pm 0.05\ \mu{\rm s}^{-1}$, 
$\nu = 2.2 \pm 0.1\ \mu{\rm s}^{-1}$. The relaxation is therefore in the regime 
of moderate motional narrowing.

\subsection{Nuclear magnetism and muon relaxation} \label{sect:Prnuc}
In this section we argue that the results of Sect.~\ref{sect:ltrelax}, together 
with the temperature dependence of the ZF relaxation rate~$\Lambda(T)$ 
(Fig.~\ref{fig:ZFtempdep}), can be understood by assuming that nuclear magnetism 
is the principal source of the $\mu^+$ local field~${\bf H}_{\rm loc}$, and that 
furthermore the $^{141}$Pr nuclear magnetism is hyperfine enhanced at low 
temperatures. In other words there is no sign of a Pr$^{3+}$ electronic magnetic 
moment, consistent with the hypothesis of a nonmagnetic CEF ground state. 

The decrease of $\Lambda(T)$ above $\sim$10 K is qualitatively consistent with 
the onset of $^{141}$Pr nuclear spin-lattice relaxation by thermally-populated 
magnetic Pr$^{3+}$ CEF excited states; the corresponding increase of $\nu$ 
motionally narrows the muon relaxation. At sufficiently high temperatures (above 
50 K, cf.\ Fig.~\ref{fig:ZFtempdep}) the exponent~$\beta$ tends to the 
val\-ue~$\beta = 2$ characteristic of a Gaussian distribution of static fields. 
There are two possible explanations of this behavior. Either the $^{141}$Pr 
relaxation remains so rapid that it is motionally narrowed and only the 
$^{115}$In nuclei contribute to the (quasistatic) muon local field, or the 
Pr$^{3+}$ fluctuations become fast enough so that their contribution to the 
$^{141}$Pr relaxation itself becomes motionally narrowed and negligible. Then 
the $^{141}$Pr nuclei relax slowly under the influence of (unenhanced) nuclear 
dipolar fields. In either case the $\mu^+$ relaxation is describable by the K-T 
model in the quasistatic limit, and one would observe essentially static nuclear 
dipolar fields at the muon sites. We shall see that the accuracy of our 
experiments does not allow us to distinguish between these two possibilities.

[It should be noted that the absence of muon relaxation by fluctating fields due 
directly to Pr$^{3+}$ CEF excitations at high temperatures is not surprising. 
These fluctuations would be expected to be rapid, at least of the order of 
$\Delta_{\rm CEF}/\hbar$, in which case arguments similar to those of 
Sect.~\ref{sect:ZFmuSR} show that the muon relaxation rate is unobservably small 
($\lesssim 0.004\ \mu{\rm s}^{-1}$).]

The above picture can be put on a more quantitative footing by comparing the 
data with the expected relaxation behavior at low and high temperatures, i.e., 
with and without hyperfine enhancement, respectively. We start by calculating 
the expected relaxation at high temperatures, where according to the above 
picture the nuclear contributions to the $\mu^+$ relaxation can be calculated 
using the usual Van Vleck method of moments. Then one has static dipolar 
broadening from $^{115}$In nuclei, together with a contribution from 
(unenhanced) $^{141}$Pr nuclei if this contribution is not motionally narrowed.

The $\mu^+$ electric charge produces an electric field gradient at the 
$^{141}$Pr ($I = 5/2$) and $^{115}$In ($I = 9/2$) sites. This produces a 
quadrupole splitting for both nuclides, which are not split in the unperturbed 
crystal because both sites possess cubic point symmetry. The quadrupole 
splitting in turn modifies the secular terms in the dipolar interaction, which 
must be taken into account in calculating the $\mu^+$ relaxation in zero 
field.\cite{HUIN79,Hart77}\, The $\mu^+$ stopping site is unknown in PrInAg$_2$. 
Calculated val\-ues of $\sigma$ for several candidate $\mu^+$ sites are shown in 
Table~\ref{tab:ZFwidths}, which gives individual 
contributions~$^{141}\sigma_{\rm ZF}$ and $^{115}\sigma_{\rm ZF}$ from 
$^{141}$Pr and $^{115}$In nuclei, respectively, together with the total 
rate~$\sigma_{\rm ZF}^{\rm tot} = \left( {^{141}\sigma_{\rm ZF}^2} + 
{^{115}\sigma_{\rm ZF}^2} \right)^{1/2}$. 
\begin{table}[t]
\begin{tabular}{ccccc}
Site & Coordinates & $^{141}\sigma_{\rm ZF}$ & $^{115}\sigma_{\rm ZF}$ & 
$\sigma_{\rm ZF}^{\rm tot}$ \\
(Wyckoff notation) & & $(\mu{\rm s}^{-1})$ & $(\mu{\rm s}^{-1})$ & $(\mu{\rm 
s}^{-1})$ \\
\hline
\noalign{\vskip2pt}
$d$ & \mbox{$(\frac{1}{4},\frac{1}{4},0)$} & 0.1329 & 0.1667 & 0.2132 \\
\noalign{\vskip2pt}
$e$ & \mbox{$(\frac{1}{4},0,0)$} & 0.2570 & 0.3227 & 0.4125 \\
\noalign{\vskip2pt}
$f$ & \mbox{$(\frac{1}{8},\frac{1}{8},\frac{1}{8})$} & 0.3906 & 0.1286 & 0.4113 
\\
\noalign{\vskip2pt}
Observed & & & & $0.18 \pm 0.02$ \\
\end{tabular}
\vspace*{12pt} \caption{Calculated powder-average ZF-$\mu$SR static Gaussian 
relaxation rates $\sigma_{\rm ZF}$ for candidate $\mu^+$ sites in PrInAg$_2$, 
assuming quadrupole splitting by the $\mu^+$ electric field gradient (Refs.~\protect\onlinecite{HUIN79} and \protect\onlinecite{Hart77}). Individual 
contributions~$^{141}\sigma_{\rm ZF}$ and $^{115}\sigma_{\rm ZF}$ from 
$^{141}$Pr and $^{115}$In nuclei, respectively, are shown, together with the 
total rate~$\sigma_{\rm ZF}^{\rm tot}$. The observed high-temperature rate is 
given for comparison.} \label{tab:ZFwidths}
\end{table}
Best agreement with the observed high-temperature static rate of $0.18 \pm 0.02\ 
\mu{\rm s}^{-1}$ is found for the \mbox{$(\frac{1}{4},\frac{1}{4},0)$} $d$ site 
(Wyckoff notation) with or without an (unenhanced) $^{141}$Pr contribution to 
$\sigma_{\rm ZF}^{\rm tot}$.

The muon may distort the lattice locally, thereby modifying the near-neighbor 
dipolar interactions primarily responsible for the relaxation. These dipolar 
interactions vary as the cube of the near-neighbor distances, the change of 
which can therefore be estimated from a comparison of calculated and measured 
$\mu^+$ static relaxation rates. This calculation yields a local dilatation of 
$5 \pm 4$\%, assuming the muon occupies the $d$ site and that the $^{141}$Pr 
contribution is present. This is comparable to the val\-ues~2--5\% found in 
copper under similar circumstances by Camani {\em et al.\/}\cite{CGRS77} and 
Luke {\em et al.\/},\cite{LBKN91} but the high-temperature data are equally 
consistent with no $^{141}$Pr contribution and/or little if any dilatation.

At 0.7 K the $^{141}$Pr contribution to the $\mu^+$ static relaxation rate will 
be increased by the factor~$1 + K$ if the $^{141}$Pr dipole moment is hyperfine 
enhanced (Sect.~\ref{sect:hfenh}). Using $a_{\rm hf} = 187.7\ 
\mbox{mole\,emu}^{-1}$ (Ref.~\onlinecite{AnDa77}) and the extrapolated 
low-temperature val\-ue of the Van Vleck susceptibility~$\chi_{\rm VV} \approx 
0.04\ \mbox{emu\,mole}^{-1}$ (Ref.~\onlinecite{YBMC96}), we obtain $K \approx 
7.5$ from Eq.~(\ref{eq:K}), in agreement with the val\-ue obtained from the 
nuclear Schottky anomaly in the low-temperature specific heat.\cite{MYBH98}\, 
This gives an expected low-temperature rate
\begin{eqnarray*} 
\sigma(\mbox{low $T$}) & = & \left[{^{141}\sigma}_{\rm ZF}^2(1 + K)^2 + 
{^{115}\sigma}_{\rm ZF}^2\right]^{1/2} \\ 
& = & 1.1419\ \mu{\rm s}^{-1} 
\end{eqnarray*}
for the $d$ site, in excellent agreement with the observed low-temperature 
val\-ue of $1.15 \pm 0.05 \ \mu{\rm s}^{-1}$ quoted above.

We now turn to the observed val\-ue~$\nu = 2.2\ \mu{\rm s}^{-1}$ of the 
low-temperature fluctuation rate (Sect.~\ref{sect:ltrelax}). The magnitude and 
temperature independence of this rate below 10 K lead us to interpret it as due 
to the like-spin coupling between (hyperfine-enhanced) $^{141}$Pr nuclear 
moments. Using the Van Vleck method of moments,\cite{Abra61} the calculated 
$^{141}$Pr zero-field rms relaxation rate (neglecting quadrupolar splitting) is 
$^{141}\sigma_{\rm ZF} = 0.7490\ \mu{\rm s}^{-1}$. This is smaller by a factor 
of $\sim$3 than the observed fluctuation rate. 

We first attempt to resolve this discrepancy by noting that the electric field 
gradient due to the $\mu^+$ charge will induce quadrupole splitting of the 
neighboring $^{141}$Pr nuclei. The effect of this splitting on the muon dynamics 
has been taken into account, but it also has an effect on the $^{141}$Pr 
relaxation. In the presence of such quadrupole splitting the like-spin 
zero-field linewidth has only been calculated for $I = 1$ and 3/2 
(Ref.~\onlinecite{Abra61}), where it is found that $\sigma_{\rm ZF}$ is 
increased by factors of 1.19 and 1.26 respectively. These are not enough to 
explain the shortfall. It is likely that the corresponding factor for $I = 5/2$ 
is larger than 1.26, but the trend does not seem to allow explanation of a factor 
of 3 by this mechanism.

We next consider the indirect exchange interaction between $^{141}$Pr nuclei 
described in Sect.~\ref{sect:hfenh}, which is mediated by the electronic 
exchange between neighboring Pr$^{3+}$ ions. We take the observed fluctuation 
rate~$\nu$ as a measure of the nuclear exchange constant~${\cal J}_{\rm 
nuc}/\hbar$, and use $K = 7.5$ and Eq.~(\ref{eq:Jnuc}) to obtain the estimate
\[ {\cal J}_{\rm el}/k_B \approx 0.19\ {\rm K} \]
for the electronic exchange constant~${\cal J}_{\rm el}$. This val\-ue is 
reasonable when compared with results in other nonmagnetic-ground-state Pr-based 
compounds: ${\cal J}_{\rm el}/k_B = 0.61\ {\rm K}$ in PrP 
(Ref.~\onlinecite{MyNa74}) and 0.39 K in PrNi$_5$.\cite{KFBM80}\, In PrInAg$_2$ 
${\cal J}_{\rm el}$ is comparable to the Kondo energy scale of 0.5--1 K, and 
should therefore be taken into account in the theory of the nonmagnetic Kondo 
effect in this system.

\subsection{Effect of muon charge on Pr{\boldmath$^{3+}$} ions}
In addition to contributing an electric field gradient at near-neighbor nuclear 
sites, the $\mu^+$ charge produces an electric field which perturbs the CEF 
splitting of Pr$^{3+}$ near neighbors. The symmetry of these Pr$^{3+}$ sites is 
lowered and some degeneracies, including that of the $\Gamma_3$ ground-state 
doublet, are lifted. The corresponding modification of the Van Vleck 
susceptibility has been observed by transverse-field $\mu$SR (TF-$\mu$SR) in a 
number of Pr-based intermetallics.\cite{FAGG95a,TAGG97}\, The ground state 
singlet is, however, still nonmagnetic, and Van Vleck paramagnetism and the 
corresponding nuclear hyperfine enhancement will remain features of the 
perturbed system. 

If the Pr$^{3+}$ ground state were magnetic ($\Gamma_4$, $\Gamma_5$), no 
qualitative effect of the $\mu^+$ charge would expected except under extreme 
conditions. If it were argued that in PrInAg$_2$ the $\mu^+$ electric field 
splits a magnetic Pr$^{3+}$ ground state so that the perturbed ground state is 
nonmagnetic, then this perturbation must split the degeneracy by an amount of 
the order of the unperturbed excitation energy ($\Delta_{\rm CEF}/k_B \sim 60$ 
K) to explain the absence of temperature dependence of the ZF relaxation rate 
below 10 K (Fig.~\ref{fig:ZFtempdep}). This would be an improbable coincidence, 
and furthermore the perturbation would have to be considerably larger than 
observed previously (see the following section).

\subsection{Comparison with {\boldmath$\mu$}SR and NMR in other Pr-based compounds} \label{sect:others}
We review some earlier results in nonmagnetic ground state Pr-based compounds 
for comparison with the present work. 

\paragraph{PrNi$_5$.} The ground state in this compound is a $\Gamma_1$ singlet, 
and the smallest CEF splitting is ${\sim}k_B{\times}23$ K.\cite{ABGG92}\, ZF- 
and TF-$\mu$SR studies\cite{FAGG95a,FAGG95b} suggest considerable perturbation 
of the Pr$^{3+}$ CEF levels by the $\mu^+$ electric field, since the TF-$\mu$SR 
frequency shift does not track the bulk susceptibility. A reduction of the CEF 
splitting by ${\sim}k_B{\times}10$ K (i.e., $\sim$100\%) is needed to account 
for the modified susceptibility. 

$^{141}$Pr NMR has also been reported in PrNi$_5$.\cite{LGSZ79,KWG80}\, 
Estimates of the hyperfine enhancement parameters ${\cal J}_{\rm el} \approx 
0.02$ eV and $K \approx 12$ have been obtained from the data. We use these to 
calculate ${\cal J}_{\rm nuc}/\hbar \approx 5.5 \times 10^6\ {\rm s}^{-1}$ from 
Eq.~(\ref{eq:Jnuc}), which is roughly consistent with an independent estimate of 
the ``zero-field splitting'' ${\lesssim} 10^6\ {\rm s}^{-1}$ from the 
low-temperature NMR data.

\paragraph{PrIn$_3$.} In this singlet-ground-state compound the Pr$^{3+}$ ions 
have considerably larger CEF splittings ($\Delta_{\rm CEF}/k_B = 101$ K; cf.\ 
Ref.~\onlinecite{BdWvD69}) than in PrNi$_5$. As in the latter compound the 
TF-$\mu$SR frequency shift does not track the bulk 
susceptibility,\cite{GFTK97,TAGG97} but the discrepancy is much smaller in 
PrIn$_3$. The change in $\Delta_{\rm CEF}/k_B$ was found to be of the same order 
of magnitude ($\sim$10 K) as in PrNi$_5$, so that the smaller effect on the 
susceptibility is related to the inverse relation between $\Delta_{\rm CEF}$ and 
$\chi_{\rm VV}$ [Eq.~(\ref{eq:chiVV})]: a given absolute change of $\Delta_{\rm 
CEF}$ has a smaller effect on $\chi_{\rm VV}$ when $\Delta_{\rm CEF}$ is large. 
Neither ZF- nor LF-$\mu$SR has been reported to date in PrIn$_3$.

$^{141}$Pr NMR in PrIn$_3$ (Ref.~\onlinecite{SKYT81}) yields $2.3 \pm 0.3\ 
\mu{\rm s}^{-1}$ for the spin-spin relaxation rate $1/T_2$, which is comparable 
to the val\-ue of $\nu$ found in PrInAg$_2$ as described above. But $1/T_2$ 
increases with increasing temperature below 4.2 K, whereas $\mu^+$ motional 
narrowing is observed only above $\sim$10 K\@. This discrepancy is not 
understood.

\section{Conclusions} \label{sect:concl}
Two features of the ZF- and LF-$\mu$SR results in PrInAg$_2$ corroborate the 
conclusion of Yatskar {\em et al.\/}\cite{YBMC96} that the Kondo behavior of the 
low-temperature specific heat in this compound originates from an unconventional 
Kondo effect associated with a nonmagnetic Pr$^{3+}$ $\Gamma_3$ CEF ground 
state. First, the fact that the data show no magnetic anomaly between 100 mK and 
10 K rules out both static magnetism and Kondo spin fluctuations associated with 
the degenerate Pr$^{3+}$ CEF ground state, so that neither of these mechanisms 
can be responsible for the low-temperature specific heat anomaly. 

Second, the low-temperature muon-spin dynamics can be quantitatively understood 
in terms of nuclear magnetism only; there is no sign of a Pr$^{3+}$ electronic 
magnetic moment. Furthermore, quantitative agreement is obtained only if the 
$^{141}$Pr nuclear magnetism is hyperfine-enhanced, which can occur only if the 
Pr$^{3+}$ CEF ground state is nonmagnetic. The experimental val\-ue of the 
$\mu^+$--$^{141}$Pr dipolar coupling is in good agreement with that calculated 
assuming hyperfine enhancement of the $^{141}$Pr nuclear dipole moment. In 
addition, the experimental estimate of the coupling between $^{141}$Pr nuclei 
is about 3 times larger than the val\-ue calculated neglecting the indirect 
nuclear coupling which arises from hyperfine enhancement. This strongly suggests 
the existence of such an indirect mechanism; the required val\-ue of the 
electronic exchange interaction between Pr$^{3+}$ ions (${\cal J}_{\rm el}/k_B 
\approx 0.19$ K) is less than but comparable to that in similar Pr-based 
compounds with nonmagnetic CEF ground states. Both these results depend 
crucially on hyperfine enhancement, and thus are evidence for the absence of a 
Pr$^{3+}$ ground-state magnetic moment. We emphasize that this electronic 
exchange should be taken into account in a complete theory of a nonmagnetic 
Kondo effect in PrInAg$_2$.

\medskip We are grateful to W.~P. Beyermann and R. Mov\-sho\-vich for discussing 
their research with us. One of us (D.E.M.) wishes to thank the staff of the 
MST-10 group, Los Alamos National Laboratory, for their hospitality during his 
stay at Los Alamos. This work was supported in part by the U.S. National Science 
Foundation, Grant Nos.~DMR-9418991 (Riverside) and DMR-9510454 (Columbia), and 
in part by the U.C. Riverside Academic Senate Committee on Research, the 
Director for Energy Research, Office of Basic Energy Sciences, U.S. Department 
of Energy (DOE) (Ames), the Japanese agency NEDO (Columbia), and the Netherlands 
agencies FOM and NWO (Leiden); and was carried out in part under the auspices of 
the U.S. DOE (Los Alamos). Ames Laboratory is operated for the U.S. DOE by Iowa 
State University under Contract No.~W-7405-Eng-82.

%\bibliographystyle{prsty}				% PR/PRL style.
%\bibliographystyle{plain}				% alphabetized, with titles.
%\bibliography{praseo,qke,musr,nmr}

\begin{thebibliography}{10}
\bibitem{Prrelax}
In the following we show that the $^{141}$Pr contribution to muon relaxation in
  PrInAg$_2$ is in fact dynamic, and thus constitutes an exception to this
  rule.
\bibitem{Prmoment}
A significant reduction of the Pr$^{3+}$ moment at 10 K and below would itself
  signal a Kondo effect with $T_K$ much larger than 10 K. This would be in
  contradiction with the observed specific heat
  (Ref.~\protect\onlinecite{YBMC96}), which agrees well with a CEF-level
  calcuation with no high-temperature Kondo effect. It would also leave
  unexplained the observed large (${\sim}0.8\ \mu{\rm s}^{-1}$) and
  temperature-independent $\mu^+$ relaxation rate below 10 K
  (Fig.~\protect\ref{fig:ZFtempdep}).
\end{thebibliography}
%\end{document}

\end{document}